# Spectroscopic study of double-walled carbon nanotubes functionalization for preparation of carbon nanotube / epoxy composites

**Vincent Leon, Romain Parret, Robert Almairac, Laurent Alvarez, Ahmed Azmi Zahab, Jean-Louis Bantignies**
*Laboratoire Charles Coulomb UMR 5221, CNRS-Université Montpellier 2, F-34095, Montpellier France*

**Moulay-Rachid Babaa**
*Chemical Engineering Department, Universiti Teknologi PETRONAS, Bandar Seri Iskandar, 31750 Tronoh, Perak, Malaysia*

**Brian P. Doyle**
*TASC-INFM Laboratory, Area Science Park – Basovizza, I-34012, Trieste, Italy*

**Patrick Ienny**
*CMGD, Ecole des Mines d'Alès, 7 rue Jules Renard, 30100 Alès, France*

**Philippe Parent**
*Laboratoire de Chimie-Physique, Matière et Rayonnement, CNRS-Université Pierre et Marie Curie, UMR 7614*

## ABSTRACT

A spectroscopic study of the amino functionalization of double-walled carbon nanotube (DWCNT) is performed. Original experimental investigations by near edge X-ray absorption fine structure spectroscopy at the C and O K-edges allow one to follow the efficiency of the chemistry during the different steps of covalent functionalization. Combined with Raman spectroscopy, the characterization gives a direct evidence of the grafting of amino-terminated molecules on the structural defects of the DWCNT external wall, whereas the internal wall does not undergo any change. Structural and mechanical investigation of the amino functionalized DWCNT / epoxy composites show coupling between epoxy molecules and the DWCNTs. Functionalization improves the interface between amino-functionalized DWCNT and the epoxy molecules. The electrical transport measurements indicate a percolating network formed only by inner metallic tubes of the DWCNTs. The activation energy of the barriers between connected metallic tubes is determined around 20 meV.

# I. INTRODUCTION

Individual carbon nanotubes (CNT) exhibit extraordinary mechanical and electronic properties [1-5]. Since Iijima's landmark paper in 1991 [6], because of their potential applications in microelectronics, many studies have been focused on the transfer of these properties to the macroscopic scale by elaboration of composite materials [7-12]. Nevertheless a major difficulty remains in the control of both nanotubes dispersion into the polymer matrix and interaction between the matrix and the nanotubes. The functionalization of the CNT surface is an interesting way to improve both the ability to disperse the nanotubes homogeneously throughout the matrix and to transfer the physical and electronic properties from the nanotubes fillers to the matrix [13-22].

High tensile strength and modulus, low shrinkage during cure, good chemical and corrosion resistance and high adhesion are among the properties making epoxy resin extensively used in several domains of industry. However, epoxy resin displays low strain-to-failure (~ 5%) and poor electrical conductivity, leading to drastic limitations in several domains such as aircraft applications [23-27].

Many studies have been made using epoxy and functionalized CNT [28-38]. The key role of nanotube functionalization in the interfacial adhesion has been demonstrated. Nevertheless, experimental proofs of the covalent functionalization of the tubes remain a difficult task. By the way, controversial results concerning the improvement of both the mechanical and the electrical properties of the composite materials are reprted [39-43]. In this context, experimental evidences of the different functionalization steps are required. A proper understanding of the fundamental physical interaction between the functionalized nanotubes and the epoxy matrix is therefore needed.

The first aim of our work is to obtain direct experimental evidences of the different steps of amino covalent functionalization of DWCNT. The second aim is to quantify the influence of this functionalization on the structural and electronic properties of amino-functionalized double-walled CNT/epoxy composites. An original approach coupling near edge X-ray absorption fine structure spectroscopy (NEXAFS) and Raman spectroscopy was used to study the functionalization of the DWNT. The structural and mechanical investigations of the composites were carried out using X-ray diffraction (XRD) and image correlation technique The temperature dependence of the electrical transport was determined in order to characterize the percolation network of functionalized nanotube loads in the composites.

# II. EXPERIMENTAL DETAILS

## II.1 Materials

*N,N'*-dimethylformamide anhydrous (DMF, 99.8%), 1-ethyl-3-(3-dimethylaminopropyl) carbodiimide hydrochloride (EDC), sulphuric acid ($H_2SO_4$, 95-98%) and nitric acid ($HNO_3$, > 69%) were purchased from Sigma-Aldrich, and 1,6-hexane diamine (HDA, > 99%) was obtained from Fluka. Composite materials were obtained using bisphenol-A and epichlorhydrin based resin (Epon 828) cured with methyltetrahydrophthalic anhydride (Aradur 917) as hardener and 1-methylimidazole (DY 070) as catalyst, all kindly supplied by Nanoledge Inc. (Montréal, Canada). Commercial Double walled carbon nanotubes (CCVD method) were kindly provided by Thomas Swan & co. Ltd. (Consett, UK).

## II.2. Functionalization process

In case of chemical functionalization of the CNT surface, creating defects can be seen as a drawback as they alter the structural properties of the nanotubes. One way to get around this problem is to use DWCNTs or multi-walled carbon nanotubes. While the external layer is functionalized, the internal layers are supposed to remain intact so that the overall properties of the carbon nanotubes are preserved. The chemical functionalization of the CNT surfaces occurs in two steps. First, pristine DWCNTs (P-DWCNT) were oxidized using a $HNO_3$ treatment. Such an operation is known to eliminate metallic particles resulting from the synthesis procedure but also to generate functional groups, in particular carboxylic groups [44,45]. Then amino groups are grafted using hexane diamine (HDA) and EDC as coupling agent (Fig. 1) following a method described elsewhere [46].

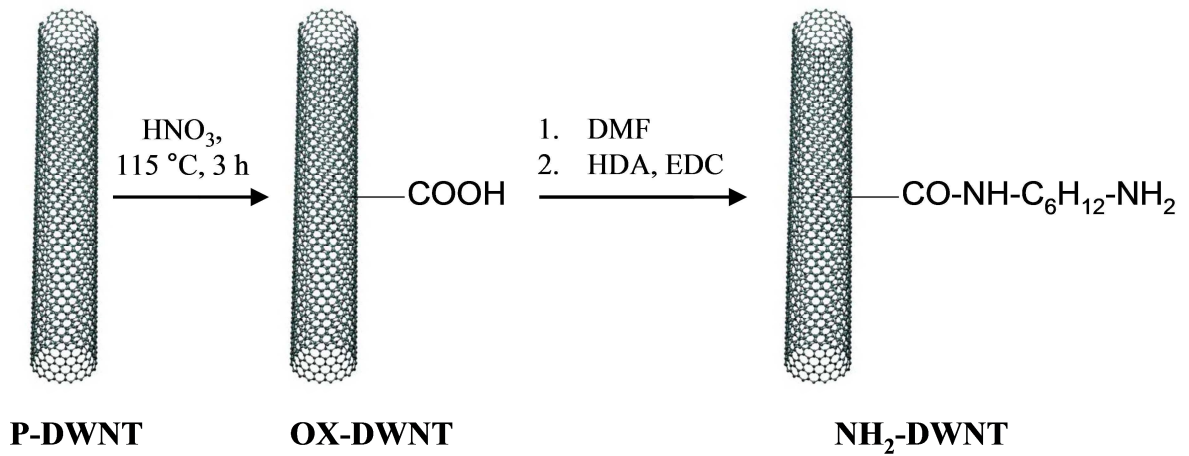

**Fig. 1.** Schematic presentation of functionalization reaction.

Carboxylic groups are generated at the surface of nanotubes through $HNO_3$ treatment, to create oxidized DWCNTs (OX-DWCNT) [47]. Typically, 350 mg of P-DWCNT were dispersed in 450 mL of $HNO_3$ in ultrasound for 45 minutes. Then the dispersion was refluxed at 115 °C for 3 h. After filtration (cellulose nitrate filter, 0.45 μm) and neutralization, OX-DWCNTs are dispersed in DMF (0.5 mg of OX-DWCNT per mL of DMF) in ultrasound (40 minutes) and vigorous stirring (1 h). Amino groups in the form of HDA are attached to the DWCNT surface using EDC as a coupling agent. EDC is used instead of DCC since it can be easily eliminated after filtration thanks to its good solubility in water. Typically, 50 mg of HDA and 50 mg of EDC are added to 250 mg of OX-DWCNT in DMF and vigorously stirred for 24 h. Amine functionalized nanotubes ($NH_2$-DWCNT) are finally obtained after filtration through PTFE filters.

## II.3. Composites elaboration

Dispersion of nanotubes in the resin is operated using mechanical strain with high shearing. Different samples of epoxy resin/carbon nanotube composite are prepared, containing from 0.20 to 0.50 wt% of $NH_2$-DWCNT with the following composition for epoxy resin: 52 wt% of resin, 47 wt% of hardener and 1 wt% of catalyst. The mixture is then degassed under vacuum. Finally samples of the composite material are heated up to 80 °C for 4 hours, then cured at 120 °C for 4 more hours. Films are obtained by compression between two steel plates coated with a PTFE film, the thermal cycle remaining unchanged. A demixing phenomenon is observed for films made with a low functionalized CNT weight fraction (below 0.25 wt%). This phase separation becomes more pronounced when the applied compression strain is greater. Films are used to measure the electrical conductivity and to investigate the structural properties (XRD, Raman spectroscopy).

## III.  EXPERIMENTAL.

### III.1.  NEXAFS

NEXAFS experiments are carried out in an ultrahigh vacuum chamber (8 x $10^{-11}$ mbar) of the BEAR beamline at the ELETTRA synchrotron radiation facility (Trieste, Italy). The X-ray spot area is 0.05 mm$^2$. The spectral resolving power (E/ΔE) of incident photons at 400 eV is 1800. Experiments at the C1s and O1s threshold are performed in the total electron yield mode. The solid powder of nanotubes is pressed on an indium metal plate. The samples are outgassed for 3-4 hours before measurements.

### III.2.  XRD

X-ray diffraction (XRD) experiments are carried out using an INEL CPS120 powder diffractometer equipped with a 120° curved detector, a Cu K$α$ x-ray source ($λ$ = 1.5418 Å) and a germanium monochromator, with a step size of 0.03° (2$θ$) and a $Q$ range from 0.14 to 7 Å$^{-1}$. The spot size is around 0.5 mm$^2$.

### III.3.  Raman spectroscopy

Room-temperature Raman spectra are recorded using the Ar/Kr laser lines at 514.5 nm ($E_L$ = 2.41 eV) in the backscattering geometry on a triple subtractive Jobin-Yvon T64000 spectrometer equipped with a liquid nitrogen cooled charge coupled device detector. The power impinging on the sample is ~ 200 $μ$W. The instrumental resolution is 2 cm$^{-1}$. The acquisition time is 300 s. For each sample, at least 4 spectra corresponding to different areas of the sample are recorded.

### III.4.  Conductivity

DC conductivity measurements are performed following the 4-points method using a Keithley Instruments Inc. voltage generator. The current is chosen in order to be in a linear region of the I-V curve. The measurements cover a temperature range from 35 to 150°C. The sample is in contact with 4 copper wires on a ceramic support allowing both electrical isolation and thermal conduction, and under vacuum (10$^{-6}$ mbar).

### III.5.  Mechanical tests

Tensile tests are performed using dumbbell-like samples (170 mm × 4 mm × 80 mm × 10 mm). An image correlation technique is used including a ZWICK TH010 traction machine (1 mm/min) and a Redlake Megaplus 2.8 bits high resolution CCD camera (1920 × 1080 pixels) with CinEMA digital image correlation software (Ecole des Mines d'Alès, France).

## IV. Results and Discussion

### IV.1. *Functionalization*

#### IV.1.1. NEXAFS

Investigation of the nature of interaction between functional groups and nanotubes at each step of the functionalization processes is performed using NEXAFS. C K-edge spectra of DWCNTs before and after the different functionalization steps are shown in Fig. 2. The spectrum measured on the pristine sample (P-DWCNT) is compared to the oxidized sample (OX-DWCNT) and nanotubes after HDA functionalization ($NH_2$-DWCNT). P-DWCNT is dominated by a sharp pre-edge $\pi^*$ peak at 286 eV associated to transition from C1s to the unoccupied $sp^2$ $\pi^*$ orbital. The broad resonances above 292 eV are associated with overlapped 1s $\rightarrow$ $\sigma^*$ transitions. The domain between 287 and 291 eV is particularly interesting in our case since it evidences the CNT chemical modifications by functionalization [44,48]. It can be noticed that in the case of OX-DWCNT, a resonance signal around 289 eV is seen. This transition is still present after functionalization with HDA ($NH_2$-DWCNT). Specific sensitivity to oxygenated functionalities attached to the surface of the nanotubes were reported; the peak at 289 eV corresponds to the C=O $\pi^*$ resonance in CNT-(OH-C*=O) or CNT*-(OH-C=O) (C* indicates the photoexcited carbon atom) [44,49]. In the case of hydroxyl groups or alcohol functions, the energy transition should be shifted at higher energy around 290 eV [50]. Therefore, these NEXAFS results show the presence of carboxylic acid groups after oxidation at the surface of the tube. Similar peaks were observed in the case of CNTs treated with strong acid baths and related to the presence carbonyl functionalities [46,48] in agreement with infrared experiments [51,52].

Fig. 3 shows the O K-edge spectra for P-DWCNT, OX-DWCNT and $NH_2$-DWCNT, respectively. In addition, spectra of acetic acid $CH_3COOH$ and acetone $CH_3COCH_3$ model molecules are also displayed. The spectra confirm the presence of oxygen containing functional groups at each step. After oxidation, the strong sharp $\pi^*$ excitation around 532 eV corresponds to carbonyl groups C=O [53]. By comparison with the O K-edge fingerprints of the carboxylic acid $CH_3COOH$ and the corresponding ketone, we are able to assign the excitation centered at 534.8 eV in OX-DWCNT to the presence of the hydroxyl groups of the carboxylic acid functional groups. The broad excitations between 538 and 550 eV are assigned to 1s $\rightarrow$ $\sigma^*$ resonances [44]. After functionalization with HDA, the disappearance of the hydroxyl fingerprint (peak at 534.8 eV) is an unambiguous sign of the attachment of HDA at the surface of the tubes. By contrast, as expected, the $\pi^*$ C=O excitation remains.

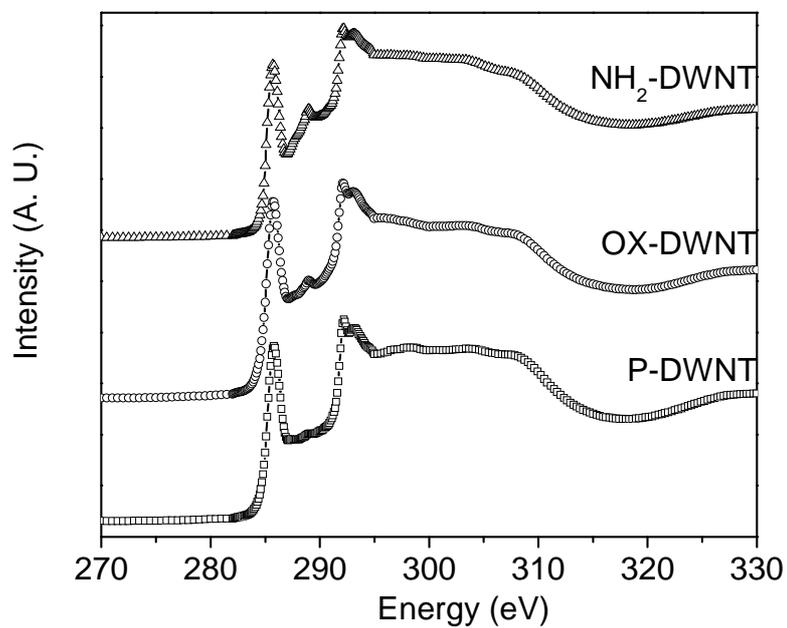

**Fig. 2.** C K-edge NEXAFS spectra of P-DWCNT, OX-DWCNT and $NH_2$-DWCNT samples.

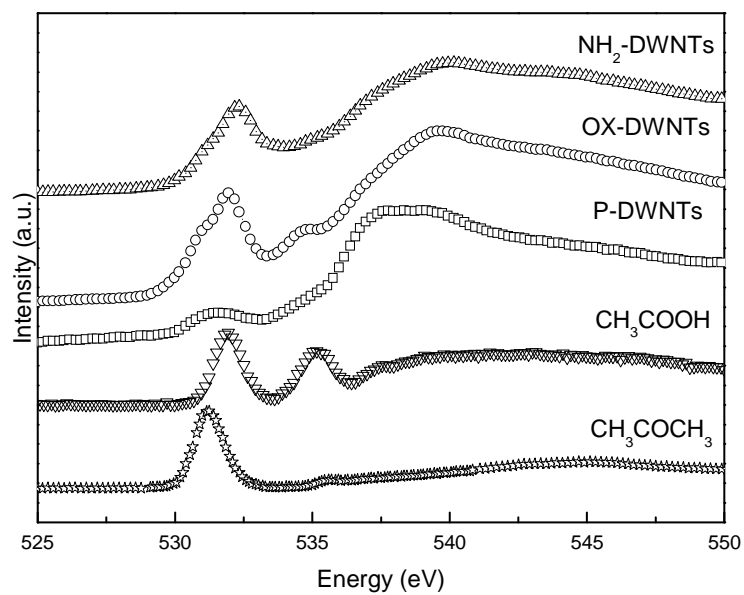

**Fig. 3.** O K-edge NEXAFS spectra of P-DWCNT, OX-MWCNT and $NH_2$-DWCNT samples. Model compounds $CH_3COCH_3$ and $CH_3COOH$ are also given for comparison.

*IV.1.2. Raman spectroscopy*

Fig. 4 displays the normalized Raman radial breathing modes (RBM) of raw (top), oxidized (middle) and functionalized DWCNTs (bottom). Fitting the data allows us to identify all the frequencies required to reproduce properly the experimental spectrum. Since each frequency $\omega_{RBM}$ corresponds to a given diameter $d_t$ via the relation $\omega_{RBM}$ (cm$^{-1}$) = A/$d_t$ (nm) + B, it is in principle possible to determine the outer diameter $d_t$ via the relation $\omega_{RBM}$ (cm$^{-1}$) = A/$d_t$ (nm) + B, it is in principle possible to determine the outer diameter of the double-walled nanotube and its corresponding inner diameter. The method is the following: we measure one frequency $\omega_{ot}$ on the Raman spectrum corresponding to an outer tube (i.e. at low frequency). Then the outer diameter $d_{ot}$ is determined from the relation $\omega_{ot}$ (cm$^{-1}$) = 217.8/$d_{ot}$ (nm) + 15.7 [54]. Considering that the distance between the outer and the inner walls is around 0.33 nm (as observed between two graphene layers in graphite or between two single-walled nanotubes organized into bundles), the inner tube diameter is $\omega_{in}$ (cm$^{-1}$) = 217.8/$d_{in}$ (nm) + 15.7, we can thus calculate the RBM frequency of the inner tube.

However, this method has to be handled very carefully. For instance, considering a peak located at $d_{ot}$=167 cm$^{-1}$ and assuming an uncertainty of about ±3cm$^{-1}$ due to the experimental resolution and the fitting procedure, the error on the outer diameter determination is ±0.03 nm. Considering an intertube distance e±0.03 nm [55], the corresponding inner tube diameter is given with an uncertainty of ±0.06 nm. The resulting uncertainty on the Raman wavenumber associated to the inner tube is then ±22 cm$^{-1}$. In addition, coupling between the walls can take place depending on the intertube distance, leading to significant shift of the Raman frequencies [55-58]. A rather good agreement between calculated and measured Raman wavenumbers can be found by using two different relationships for the outer and the inner tubes respectively depending on the intertube coupling. This method is justified by differences in the environment around both tubes as reported by D. Levshov et al. [58]. Here, our method is only used to discriminate between outer and inner tubes.

Consequently, we use this method to get some structural information on the double-walled nanotubes. RBM frequencies corresponding to outer and inner tubes are identified by the same number on figure 4 for the most intense peaks. It is sometimes not possible to find out the two walls of a DWNT (for instance the peak at 184 cm$^{-1}$ has no internal tube counterpart, expected around 350 cm$^{-1}$). This means either that the nanotube is single-walled or that one of the wall (outer or inner) is out of resonance. RBM below 195 cm$^{-1}$ correspond to outer tubes whereas the peaks above are assigned to inner tube [59]. According to the Kataura plot, with this excitation wavelength (514.5 nm), all the DWCNTs have a semiconducting outer tube, whereas the inner tubes are mainly metallic. As shown in Fig. 4, the oxidization step substantially affects the Raman spectra in the low frequency range. Indeed, peak intensities corresponding to the outer tubes (left part) are significantly reduced after oxidization with respect to the inner tubes (right part). Consequently, as expected, functionalization takes place mostly at the surface of the outer tubes, leaving the inner ones either unaffected or only slightly altered. The modifications of the tubes are also observed in the graphite-like modes region where the profile significantly changes after oxidization and even after amino group grafting (Fig. 5). We can first notice an increase of the D-band intensity, usually assigned to the presence of defects (around 1345 cm$^{-1}$), just after the oxidization step. The intensity ration between the D and the G bands increases from 11 to 28%. Such a behavior is expected as the oxidization step strongly alters the nanotubes surface by changing carbon sp$^2$ bonding into sp$^3$. Surprisingly, a D-band upshift (from 1347 to 1358 cm$^{-1}$) takes place after amino group grafting, concomitant with the appearance of the D* band at 1620 cm$^{-1}$. Those modifications reveal significant structural changes surprisingly induced by the NH$_2$ group grafting. The features around 1564 and 1594 cm$^{-1}$ are assigned respectively to G$^-$ and G$^+$ modes. The G$^-$ band profile is also strongly affected by both the oxidization and the amino group grafting. This result is expected regarding the position dependence of this mode with the tube diameter. The G$^-$ band modification is thus consistent with the vanishing of some RBM corresponding to outer tubes. A concomitant downshift of the $G^+$ mode from 1594 to 1592 cm$^{-1}$ is also observed, probably due to electronic property modifications induced by the oxidization and the grafting steps. Those behaviors of the D and the $G^-$ bands have already been observed in carbon nanotubes under electron beam irradiation [60]. We assume that the oxidization step occurring before the functionalization step creates a great number of defects on the outer wall surface in agreement with previous studies [61], leading to strong structural modifications that can explain our experimental Raman spectra.

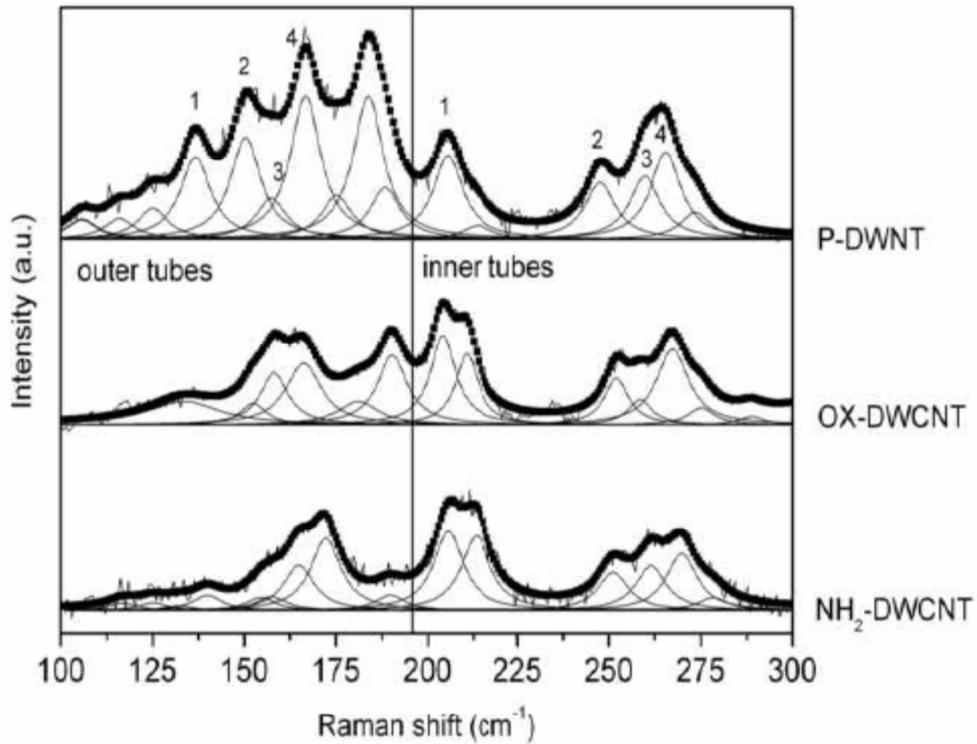

**Fig. 4.** RBM modes and the subsequent fits of P-DWCNT sample (top) and NH$_2$-DWCNT sample (bottom) at $\lambda = 514.5$ nm. Corresponding outer and inner tubes are labeled with the same number. The vertical line shows the frequency domain limit for outer and inner tubes.

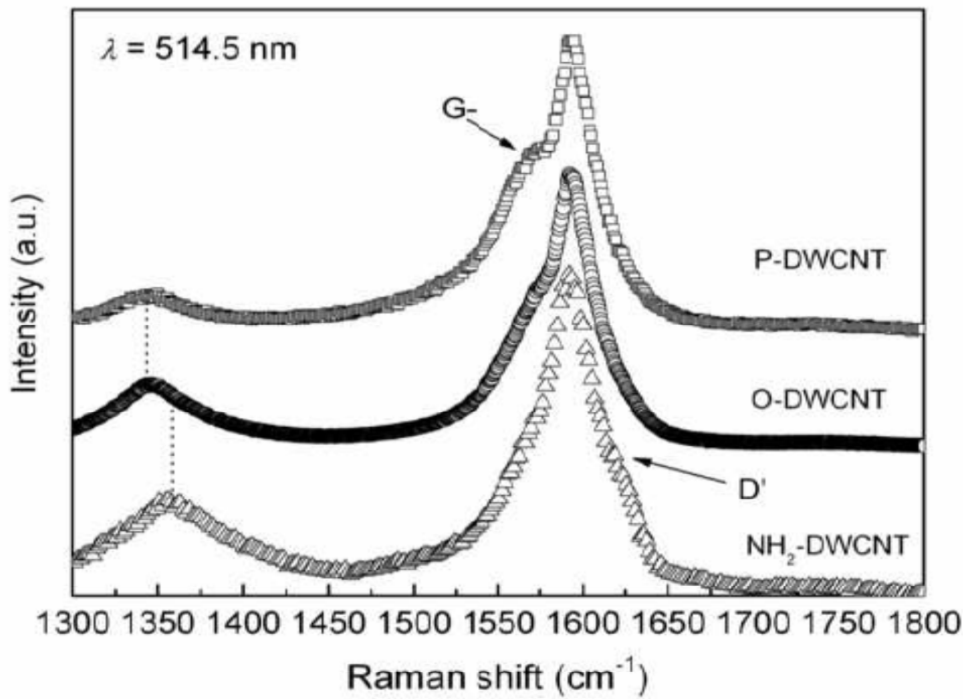

**Fig. 5.** Tangential modes of P-DWCNT and NH$_2$-DWCNT samples.

### IV.2  Mechanical tests

The image correlation technique is used to determine the displacement field at the surface of the material when strained. Results are obtained for pure epoxy, epoxy with respectively 0.25 and 0.5 wt % of P-DWNTs, and epoxy with respectively 0.25 and 0.5 wt % of NH$_2$-DWNTs. Young's modulus (3.6 ±0.05 GPa for epoxy) increases from 10 % to 15 % for a loading of 0.25 wt % of P-DWNTs (3.95 ±0.05 GPa) and NH$_2$-DWNTs (4.15 ±0.05 GPa) respectively. The functionalization process improves moderately the interface between load and matrix. As a consequence, the mechanical properties improve by 5%. However Young's modulus is 10 % lower for a 0.5 wt % loading compared to 0.25 wt %. This decrease for higher quantity of load can be explained considering a less efficient dispersion of nanotubes, due to the formation of aggregates.

### IV.3  Composite structure

The configuration of the epoxy molecules in the composite system has been investigated using XRD. The two peaks observed around 0.5 Å$^{-1}$ and 1.2 Å$^{-1}$ (Fig. 6) are the signatures of two different distances in the epoxy composite, namely ~12.5 Å and 5.5 Å. Since these distances are large compared to interatomic ones they can be assimilated to typical distances in an epoxy molecule or a fragment of the molecule which can be represented by a segment of a straight line with a uniform density of scatterer. The epoxy network can be viewed as made of connected segments of ~20 Å, some connections being realized by the hardener molecule of ~5.5 Å at the extremity of the segment. With this assumption, the diffraction pattern displays a peak at $Q$ ~1.2 Å$^{-1}$. The peak at 0.5 Å$^{-1}$ could be obtained by using the same object with a distance of 12.5 Å instead of 5.5 Å. But 12.5 Å is not the size of the hardener molecule and such a hypothesis is not valid according to the molecular configuration. To obtain this peak one must consider more complicated patterns formed with the elementary segments. A possible configuration is proposed in Fig. 6 (sketch (a) bottom object). Here the epoxy molecule has a "boomerang-like" configuration with a 33° angle between two segments of 20 Å, the distance of the link corresponding to hardener molecule is 5.5 Å.

The introduction of CNT fillers slightly but clearly shifts the low $Q$ peak towards lower wave vectors of 4 to 5%, whereas the peak for higher $Q$ values is shifted to a higher wave vector (around 1.5%) (Fig. 6, inset). CNTs are very large compared to epoxy molecules. The functionalization effect is to decrease the shorter distance (5.5 Å) and to increase the larger one (12.5 Å). A structural modification of the diffraction pattern induced by the CNT load is exhibited. According to mechanical tests, it proves that coupling between epoxy molecules and CNT does occur. A possible sketch of the epoxy in interaction with the tube is proposed on figure 6.

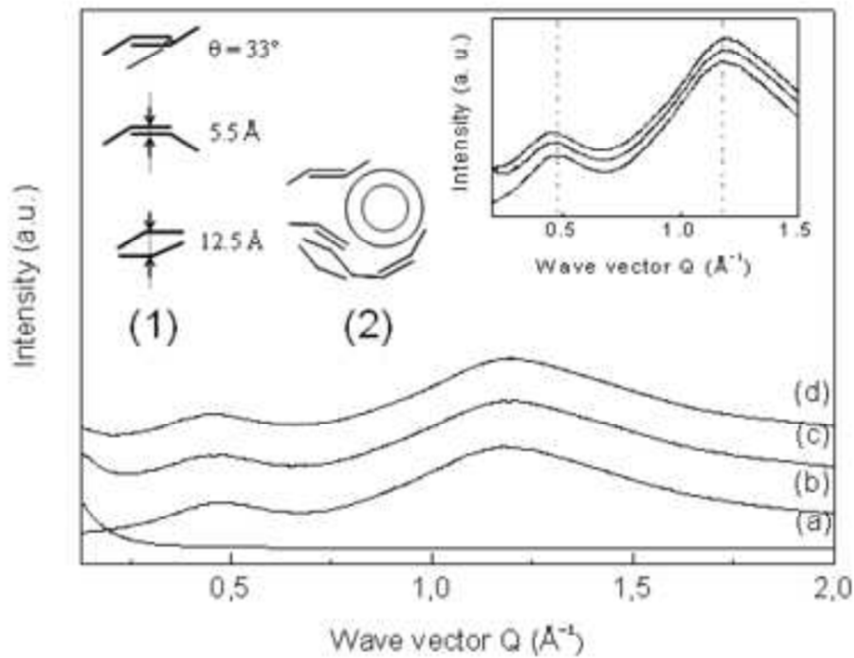

**Fig. 6.** From bottom to top: experimental diagrams of (a) P-DWCNT sample, (b) pure epoxy, (c) epoxy/0.5% P-DWCNT and (d) epoxy/0.5% NH$_2$-DWCNT. Inset represents an enlarged view of the diagram showing the shift of $Q$ values (peak around 0.5 Å$^{-1}$) after the introduction of CNT fillers into epoxy. Sketches of the epoxy structural arrangement (1) and for epoxy in interaction with the DWNTs (2)

### IV.4. Transport measurements

CNT fillers in all composite materials act as a percolating network with an ohmic behavior (Fig. 7, inset). By measuring conductivity as a function of the CNT weight fraction, we can clearly see that the value for the composite with 0.25 wt% of CNTs, corresponding to the percolation threshold, is more than 5 orders of magnitude higher (see Table 1) than the one of pure epoxy resin ($< 10^{-11}$ S/m) which is in agreement with classical percolation threshold obtained for nanotube-based composites [62]. The key point for such a low percolation threshold is the dispersion of CNTs and their high aspect ratio. An increase of electrical conductivity as a function of CNT weight fraction from around $10^{-4}$ to $3.8\ 10^{-4}$ S/m is observed at 40 °C, going up to $4.8\ 10^{-4}$ S/m at 120 °C (Table 1). To obtain the activation energy of the conductive process, we plot ln$R$ as a function of $1/T$, where $R$ and $T$ are resistance and temperature respectively. As shown in Fig. 7, the resulting plots for each sample studied are straight lines. This is a typical semi-conducting behavior for composites, following an Arrhenius-like law. The activation energy (AE) can thus be derived from the slope of each corresponding line. AEs are shown for all samples below 0.3 eV. The values found for activation energies are respectively AE = 16 meV for 0.20 wt % P-DWCNT/epoxy, AE = 24 meV for 0.50 wt % NH$_2$-DWCNT/epoxy and AE = 24 meV for 0.50 wt % P-DWCNT/epoxy. As semiconducting CNTs are not conductive in this range of temperature (the gap is around 1eV), we conclude that the percolation network is formed only by crossed metallic tubes and AE represents the potential barrier between two connected tubes. AE values obtained for the different samples show a correlation between AE and CNT loading. An increase of 66% is exhibited when the concentration is doubled. Thus barrier width changes with the CNT weight fraction, a possible explanation being the increase of the mechanical constraints at the interface between tubes in the composites as emphasized by diffraction and mechanical tests. It is also clear from the plots that the amino-covalent functionalization of the outer damaged tubes does not affect the electronic transport properties of such DWCNT based composites. This result shows that conductivity comes from the inner metallic tubes of crossed DWCNTs in the composite.

**Table 1.** Conductivity of epoxy/nanotubes composite material as a function of the carbon nanotube weight fraction.

| Carbon nanotubes weight fraction (%) | Conductivity (S.m$^{-1}$) at 40°C | Conductivity (S.m$^{-1}$) at 120°C |
|---|---|---|
| 0 | $<1.00\ 10^{-11}$ | $<1.00\ 10^{-11}$ |
| 0.25 | $8.87\ 10^{-5}$ | $1.14\ 10^{-4}$ |
| 0.5 | $3.80\ 10^{-4}$ | $4.80\ 10^{-4}$ |

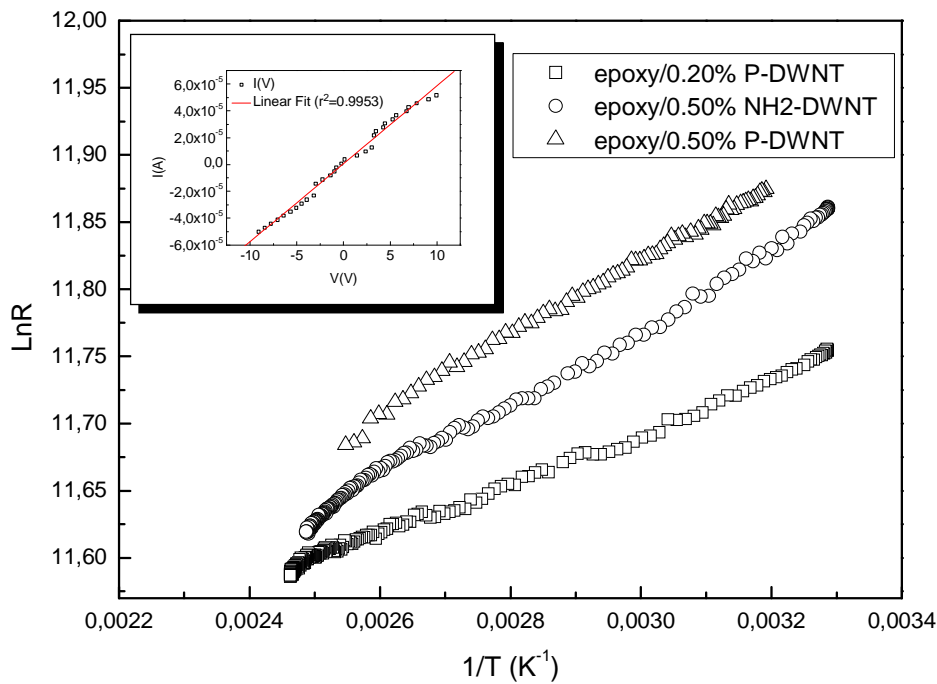

**Fig. 7.** Natural logarithm of resistance $R$ as a function of inverse temperature $1/T$ (in K$^{-1}$) for the different samples. The inset shows the ohmic behavior (linear dependence of the experimental current $I$ as a function of the voltage $V$ and linear fit of I(V)) of the epoxy/0.50 wt% P-DWCNT composite material.

## V. CONCLUSION

The effect of the covalent amino-functionalization at the surface of the tube was investigated in details. The different steps of covalent functionalization of the external tube of DWCNT by amino-terminated molecules were demonstrated using NEXAFS and Raman spectroscopy. The structure of the inner tube is preserved in this case. The structural investigation of the composites by X-ray diffraction and mechanical tests show the coupling between epoxy molecules and DWCNT. Functionalization improves the interface between NH$_2$-DWCNT fillers and the epoxy resin. The covalent functionalization of the outer damaged tubes of DWCNTs does not affect significantly the electronic transport properties of the DWCNT-epoxy composites. The electrical conductivity is controlled by inner metallic tubes of crossed DWCNTs in the composite. Finally, the potential barrier between two connected tubes has been quantified (around 20 meV).

## ACKNOWLEDGMENT

Part of this work was performed at the Synchrotron ELETTRA, Italy on BEAR beamline. This project was sponsored by European Community under Framework Program 6, Carbon Nanotubes for Application in Electronics, Catalysis, Composites and Nano-Biology (CANAPE), Contract No. NMP-500096. Nanoledge Inc. is gratefully acknowledged for supplying epoxy material.

**REFERENCES**

[1] Tracy MMJ, Ebbesen TW, Gibson JM. Exceptionally high Young's modulus observed for individual carbon nanotubes. *Nature* **1996**;381(6584):678-80.

[2] Poncharal P, Wang ZL, Ugarte D, de Heer WA. Electrostatic deflections and electromechanical resonances of carbon nanotubes. *Science* **1999**;283(5407):1513-6.

[3] Tans SJ, Verschueren ARM, Dekker C. Room-temperature transistor based on a single carbon nanotube. *Nature* **1998**;393(6680):49-52.

[4] Dai H, Wong EW, Lieber CM. Probing electrical transport in nanomaterials: conductivity of individual carbon nanotubes. *Science* **1996**;72(5261):523-6.

[5] Dresselhaus MS, Dresselhaus G, Saito R. Physics of carbon nanotubes. *Carbon* **1995**;33(7):883-91.

[6] Iijima S. Helical microtubules of graphitic carbon. *Nature* **1991**;354(6348):56-8.

[7] Li C, Thostenson ET, Chou TW. Sensors and actuators based on carbon nanotubes and their composites: a review. *Compos. Sci. Technol.* **2008**;68(6):1227-49.

[8] Shaffer MSP, Windle AH. Fabrication and characterization of carbon nanotube/poly(vinyl alcohol) composites. *Adv. Mater.* **1999**;11(11):937-41.

[9] Ajayan PM, Stephan O, Colliex C, Trauth D. Aligned carbon nanotube arrays formed by cutting a polymer resin—nanotube composite. *Science* **1994**;265(5176):1212-4.

[10] Coleman JN, Khan U, Blau WJ, Gun'ko YK. Small but strong: a review of the mechanical properties of carbon nanotube–polymer composites. *Carbon* **2006**;44(9):1624-52.

[11] Guadagno L, Vertuccio L, Sorrentino A, Raimondo M, Naddeo C, Vittoria V, et al. Mechanical and barrier properties of epoxy resin filled with multi-walled carbon nanotubes. *Carbon* **2009**;47(10):2419-30.

[12] Bachtold A, Hadley P, Nakanishi T, Dekker C. Logic circuits with carbon nanotube transistors. *Science* **2001**;294(5545):1317-20.

[13] Dyke CA, Tour JM. Covalent functionalization of single-walled carbon nanotubes for materials applications. *J. Phys. Chem. A* **2004**;108(51):11151-9.

[14] Vigolo B, Mamane V, Valsaque F, Le TNH, Thabit J, Ghanbaja J, et al. Evidence of sidewall covalent functionalization of single-walled carbon nanotubes and its advantages for composite processing. *Carbon* **2009**;47(2):411-9.

[15] Lafuente E, Callejas MA, Sainz R, Benito AM, Maser WK, Sanjuán ML, et al. The influence of single-waled carbon nanotubes functionalization on the electronic properties of their polyaniline composites. *Carbon* **2008**;46(14):1909-17.

[16] Rosca ID, Barsan MM, Butler IS, Kozinski JA. Doubly functionalized multiwall carbon nanotubes with enhanced solubility. *Carbon* **2009**;47(10):2528-55.

[17] Moniruzzaman M, Winey KI. Polymer nanocomposites containing carbon nanotubes. *Macromolecules* **2006**;39(16):5194-205.

[18]     Mun SJ, Jung YM, Kim JC, Chang JH. Poly(ethylene terephthalate) nanocomposite fibers with functionalized multiwalled carbon nanotubes via in-situ polymerization. *J. Appl. Polym. Sci.* **2008**;109(1):638-46.

[19]     Mitchell CA, Bahr JL, Arepalli S, Tour JM, Krishnamoorti R. Dispersion of functionalized carbon nanotubes in polystyrene. *Macromolecules* **2002**;35(23):8825-30.

[20]     Buffa F, Abraham GA, Grady BP, Resasco D. Effect of nanotube functionalization on the properties of single-walled carbon nanotube/polyurethane composites. *J. Polym. Sci. Part B Polym. Phys.* **2007**;45(4):490-501.

[21]     Shofner ML, Khabashesku VN, Barrera EV. Processing and mechanical properties of fluorinated single-wall carbon nanotube–polyethylene composites. *Chem. Mater.* **2006**;18(4):906-13.

[22]     Balasubramanian, K.; Burghard, M. Chemically Functionalized Carbon Nanotubes. *Small* **2005**;1(2):180-92.

[23]     Ng CB, Schadler LS, Siegel RW. Synthesis and mechanical properties of $TiO_2$-epoxy nanocomposites. *Nanostruct. Mater.* **1999**;12(1-4):507-10.

[24]     Barrau S, Demont P, Peigney A, Laurent C, Lacabanne C. DC and AC conductivity of carbon nanotubes–polyepoxy composites. *Macromolecules* **2003**;36(26):5187-94.

[25]     Moisala A, Li Q, Kinloch IA, Windle AH. Thermal and electrical conductivity of single- and multi-walled carbon nanotube-epoxy composites. *Compos. Sci. Technol.* **2006**;66(10):1285-8.

[26]     Sandler J, Schaffer MSP, Prasse T, Bauhofer W, Schulte K, Windle AH. Development of a dispersion process for carbon nanotubes in an epoxy matrix and the resulting electrical properties. *Polymer* **1999**;40(21):5967-71.

[27]     Li N, Huang Y, Du F, He X, Lin X, Gao H, et al. Electromagnetic Interference (EMI) Shielding of single-walled carbon nanotube epoxy composites. *Nano Lett.* **2006**;6(6):1141-5.

[28]     Ma PC, Mo SY, Tang BZ, Kim JK. Dispersion, interfacial interaction and re-agglomeration of functionalized carbon nanotubes in epoxy composites. *Carbon* **2010**;48(6):1824-34.

[29]     Song YS, Youn JR. Influence of dispersion states of carbon nanotubes on physical properties of epoxy nanocomposites. *Carbon* **2005**;43(7):1378-85.

[30]     Zhao JX, Ding YH. Chemical functionalization of single-walled carbon nanotubes (SWNTs) by aryl groups: a density functional theory study. *J. Phys. Chem. C* **2008**;112(34):13141-9.

[31]     Tseng CH, Wang CC, Chen CY. Functionalizing carbon nanotubes by plasma modification for the preparation of covalent-integrated epoxy composites. *Chem. Mater.* **2007**;19(2):308-15.

[32]     Yang WR, Thordarson P, Gooding JJ, Ringer SP, Braet F. Carbon nanotubes for biological and biomedical applications. *Nanotechnology* **2007**;18(41):412001.

[33]     Ge JJ, Zhang D, Li Q, Hou H, Graham MJ, Dai L, et al. Multiwalled carbon nanotubes with chemically grafted polyetherimides. *J. Am. Chem. Soc.* **2005**;127(28):9984-5.

[34]     Qin S, Qin D, Ford WT, Resasco DE, Herrera JE. Functionalization of single-walled carbon nanotubes with polystyrene via grafting to and grafting from methods. *Macromolecules* **2004**;37(3):752-7.

[35] Bahr JL, Tour JM. Covalent chemistry of single-wall carbon nanotubes. *J. Mater. Chem.* **2002**;12(7):1952-8.

[36] Huang W, Lin Y, Traylor S, Gaillard J, Rao AM, Sun YP. Sonication-assisted functionalization and solubilization of carbon nanotubes. *Nano Lett.* **2002**;2(3):231-4.

[37] Sandler JKW, Kirk JE, Kinloch IA, Schaffer MSP, Windle AH. Ultra-low electrical percolation threshold in carbon-nanotube-epoxy composites. *Polymer* **2003**;44(19):5893-9.

[38] Breton Y, Salvetat J-P, Desarmot G, Delpeux S, Sinturel C, Beguin F, et al. Mechanical properties of multiwall carbon nanotube/epoxy composites: influence of network morphology. *Carbon* **2004**;42(5-6):1027-30.

[39] Cheng QF, Wang JP, Wen JJ, Liu CH, Jiang KL, Li QQ, et al. Carbon nanotube/epoxy composites fabricated by resin transfer molding. *Carbon* **2010**;48(1):260-6.

[40] Feng QP, Yang JP, Fu SY, Mai YW. Synthesis of carbon nanotube/epoxy composite films with a high nanotubes loading by a mixed-curing-agent assisted layer-by-layer method and their electrical conductivity. *Carbon* **2010**;48(7):2057-62.

[41] Yeh MK, Hsieh TH, Tai NH. Fabrication and mechanical properties of multi-walled carbon nanotube/epoxy nanocomposites. *Mater. Sci. Eng. A* **2008**;483-484:289-92.

[42] Sun L, Warren GL, O'Reilly JY, Everett, WN, Lee SM, Davis D. et al. Mechanical properties of surface-functionalized SWCNT/epoxy composites. *Carbon* **2008**;46(2):320-8.

[43] Wang S, Liang Z, Liu T, Wang B, Zhang C. Effective amino-functionalization of carbon nanotubes for reinforcing epoxy polymer composites. *Nanotechnology* **2006**;17(6):1551-7.

[44] Kuznetsova A, Mawhinney DB, Naumenko V, Yates JT, Liu J, Smalley RE. Enhancement of adsorption inside of single-walled nanotubes: opening the entry ports. *Chem. Phys. Lett.* **2000**;321(3-4):292-6.

[45] Vigolo B, Héold C, Marêché JF, Ghanbaja J, Gulas M, Le Normand F, et al. A comprehensive scenario for commonly used purification procedures of arc-discharge as-produced single-walled carbon nanotubes. *Carbon* **2010**;48(4):949-63.

[46] Babaa MR, Bantignies JL, Alvarez L, Parent P, Le Normand F, Gulas M, et al. NEXAFS study of multiwalled carbon nanotubes functionalization with sulfonated poly(ether ether ketone) chains. *J. Nanosci. Nanotechnol.* **2007**;7(10):3463-7.

[47] Chen J, Rao AM, Lyuksyutov S, Itkis ME, Hamon MA, Hu H, et al. Dissolution of full-length single-walled carbon nanotubes. *J. Phys. Chem. B* **2001**;105(13):2525-8.

[48] Larciprete R, Gardonio S, Petaccia L, Lizzit S. Atomic oxygen functionalization of double walled C nanotubes. *Carbon* **2009**;47(11):2579-89.

[49] Braun A, Huggins FE, Shah N, Chen Y, Wirick S, Mun SB, et al. Advantages of soft X-ray absorption over TEM-EELS for solid carbon studies – a comparative study on diesel soot with EELS and NEXAFS. *Carbon* **2005**;43(1):117-24.

[50] Kuznetsova A, Popova I., Yates JT, Bronikowski MJ, Huffman CB, Liu J, Smalley RE, Hwu HH, Chen JG. Oxygen-containing functional groups on single-wall carbon nanotubes: NEXAFS and vibrational spectroscopic studies. *J. Am. Chem. Soc.* **2001**, 123 (43) :10699-704.

[51]     Sbai K, Rahmani A, Chadli H, Bantignies JL, Hermet P, Sauvajol JL. Infrared spectroscopy of multi-walled carbon nanotubes. *J. Phys. Chem. B* **2006**;110(25):12388-93.

[52]     Bantignies JL, Sauvajol JL, Rahmani A, Flahaut E. Infrared-active phonons in carbon nanotubes. *Phys. Rev. B* **2006**;74(19): 195425.

[53]     Banerjee S, Hemraj-Benny T, Balasubramanian M, Ficher DA, Misewich JA, Wong SS. *S*urface chemistry and structure of purified, ozonized, multiwalled carbon nanotubes probed by NEXAFS and vibrational spectroscopies. *ChemPhysChem* **2004**;5(9):1416-22.

[54]     Jorio A, Araujo PT, Doorn SK, Maruyama S, Chacham H, Pimenta MA. The Kataura plot over broad energy and diameter ranges. *Phys. Stat. Solidi B* **2006**;243(13):3117-21.

[55]     Han S, Goddard WA. Coupling of Raman breathing modes in double-wall carbon nanotubes and bundles of nanotubes. *J. Phys. Chem. B* **2009**; 113:7199-204.

[56]     Pfeiffer R, Simon F, Kuzmany H, Popov VN, Zolyomi V, Kurty J. Tube-Tube interaction in double-wall carbon nanotubes. *Phys. Stat. Solidi B* **2006**;243(13):3268-72.

[57]     Popov VN, Henrard L. Breathing like phonon modes of multiwalled carbon nanotubes. *Phys. Rev. B* **2002**;65:235415.

[58]     Levshov D, Than TX, Arenal R, Popov VN, Parret R, Paillet M, JourdainV, Zahab AA, Michel T, Yuzyuk Y, Sauvajol JL. Experimental evidence of a mechanical coupling between layers in an individual double-walled carbon nanotube. *Nano Lett.* **2011**;11(11): 4800-4.

[59]     Kataura H, Kumazawa Y, Maniwa Y, Umezu I, Suzuki S, Ohtsuka Y, et al. Optical properties of single-wall carbon nanotubes. *Synth. Met.* **1999**;103(1-3):2555-8.

[60]     Gupta S, Patel RJ. Changes in the vibrational modes of carbon nanotubes induced by electronbeam irradiation: resonance Raman spectroscopy. *J. Raman Spectrosc.* **2007**;38(2):188-99.

[61]     Monthioux M, Smith BW, Burteaux B, Claye A, Fischer JE, Luzzi DE. Sensitivity of single-wall carbon nanotubes to chemical processing: an electron microscopy investigation. *Carbon* **2001**;39(8):1251-72.

[62]     Gojny FH, Wichmann MHG, Fiedler B, Kinloch IA, Bauhofer W, Windle AH, et al. Evaluation and identification of electrical and thermal conduction mechanisms in carbon nanotube/epoxy composites. *Polymer* **2006**;47(6):2036-45.